\documentclass[sigconf]{acmart}

\usepackage[english]{babel}
\usepackage{colortbl}
\usepackage{acronym}
\usepackage[acronym,nomain,nonumberlist]{glossaries}

\usepackage{amsmath}
\usepackage{soul}
\usepackage{adjustbox}
\usepackage{graphicx}
\usepackage{pifont}
\newcommand{\cmark}{\ding{51}}%
\newcommand{\xmark}{\ding{55}}%

\newcommand{\mycomment}[1]{} 

\AtBeginDocument{%
  \providecommand\BibTeX{{%
    \normalfont B\kern-0.5em{\scshape i\kern-0.25em b}\kern-0.8em\TeX}}}

\setcopyright{acmcopyright}
\copyrightyear{2023}
\acmYear{2023}
\acmDOI{XXXXXXX.XXXXXXX}

\acmConference[Trustbus '23]{Make sure to enter the correct
  conference title from your rights confirmation emai}{August 29-- Sept. 1,
  2023}{Benevento, Italy}
\begin{document}

\title{Bypassing antivirus detection: old-school malware, new tricks}


\author{Efstratios Chatzoglou}
\affiliation{%
  \institution{University of the Aegean}
  \city{Karlovasi}
  \country{Greece}}
\email{efchatzoglou@aegean.gr}

\author{Georgios Karopoulos}
\affiliation{%
  \institution{European Commission, Joint Research Centre (JRC)}
  \city{Ispra}
  \country{Italy}}
\email{Georgios.KAROPOULOS@ec.europa.eu}
\authornote{Corresponding author}

\author{Georgios Kambourakis}
\affiliation{%
  \institution{University of the Aegean}
  \city{Karlovasi}
  \country{Greece}}
\email{gkamb@aegean.gr}

\author{Zisis Tsiatsikas}
\affiliation{%
  \institution{University of the Aegean}
  \city{Karlovasi}
  \country{Greece}}
\email{tzisis@aegean.gr}

\renewcommand{\shortauthors}{Chatzoglou, et al.}

\begin{abstract}
  Being on a mushrooming spree since at least 2013, malware can take a large toll on any system. In a perpetual cat-and-mouse chase with defenders, malware writers constantly conjure new methods to hide their code so as to evade detection by security products. In this context, focusing on the MS Windows platform, this work contributes a comprehensive empirical evaluation regarding the detection capacity of popular, off-the-shelf antivirus and endpoint detection and response engines when facing legacy malware obfuscated via more or less uncommon but publicly known methods. Our experiments exploit a blend of seven traditional AV evasion techniques in 16 executables built in C++, Go, and Rust. Furthermore, we conduct an incipient study regarding the ability of the ChatGPT chatbot in assisting threat actors to produce ready-to-use malware. The derived results in terms of detection rate are highly unexpected: approximately half of the 12 tested AV engines were able to detect less than half of the malware variants, four AVs exactly half of the variants, while only two of the rest detected all but one of the variants. 
\end{abstract}

\ccsdesc[500]{Security and privacy~Software security engineering}

\keywords{Malware, Antivirus software, Malware evasion techniques, ChatGPT}


\maketitle

\section{Introduction}
\label{S:Introduction}


Over the last three decades, the endless struggle between evildoers and defenders has become an ongoing cat-and-mouse game. Malicious software, commonly referred to as malware, is a low-hanging fruit for threat actors due to the ever-increasing variety of applications and services offered in cyberspace. When combined with legacy social engineering techniques, like phishing, malware is one of the most effective means of inflicting damage on the target computer system. Indeed, according to Statista, during 2022, the worldwide number of malware attacks reached 5.5 billion, an increase of 2\% compared to 2021~\cite{Statista:Malware:2022}.

Malware is often disguised as a legitimate application, making it easier to deceive the end-user in executing it~\cite{8323959}. Moreover, malware is a key factor in the creation of botnets, which are often exploited to launch large-scale catastrophic distributed denial of service (DDoS) attacks~\cite{8170867}. Each bot is a piece of malware that receives orders from a master via a command and control (C2) infrastructure. Code obfuscation is one of the several techniques used by malware to elude static analysis methods and legacy anti-malware solutions. Typically, detection is done by comparing the hash of the considered file with the known malware hashes stored in a database.

AV software is the commonest defense against malware. As already pointed out, legacy AVs use signature-based detection to identify known malware, and heuristics-based detection to perceive previously unseen malware based on its behavior. Moreover, defenders are increasingly employing advanced approaches, such as machine learning (ML), to detect and safeguard against new malware. That is, traditional (shallow) or deep learning techniques are exploited to train malware detection models, which are then used to detect previously unseen malware families and polymorphic strains~\cite{electronics9101684, Anderson2017EvadingML}.

On the other side of the spectrum, to evade both traditional and signatureless malware detection, aggressors use assorted techniques, including code obfuscation, polymorphism, and packing. Code obfuscation renders malware code difficult to read and understand by humans and machines. Moreover, polymorphism is used to create malware variants that can alter their signature or behavior to deceive the underlying detection mechanisms. Last but not least, packing is used to compress and encrypt malware code to make it cumbersome for antivirus software to analyze. More recently, adversarial ML has demonstrated that signatureless malware ML models are susceptible to gradient-based and other sophisticated attacks~\cite{Anderson2017EvadingML}. That is, through specific techniques, including generative adversarial network (GAN), the attacker's generator model learns to produce evasive variants that appear to be drawn from the benign class. 

Until now, several researches~\cite{Afianian2019, ntanto2019, You2010, smith2009computer} have assessed AV solutions with regard to their efficiency in detecting malware. Nevertheless, as also demonstrated by the work at hand, despite the substantial progress, defensive solutions, including AV and endpoint detection and response (EDR), seem to present shortcomings, even when it comes to the detection of old-school malware, which in turn questions their effectiveness. In this context, the present work provides a comprehensive assessment of the ability of well-known AVs and EDRs when dealing with legacy malware code, which however is obfuscated via more or less uncommon but publicly available methods. Particularly, the main contributions of this work vis-\`a-vis the relevant literature can be summarized as follows.

\textit{\textbf{Our contribution:}} We meticulously evaluate the detection performance of 16 commonly accepted and popular products, twelve AVs and four EDRs, when copying with obfuscated legacy malware. Obfuscation is done through custom-made, but publicly known and easily accessible techniques. Particularly, the conducted analysis involves the injection of a mix of seven traditional AV evasion techniques in 16 executables, developed in three popular programming languages, namely C++, Go, and Rust. On top of that, we contribute a preliminary study regarding the capability of the ChatGPT chatbot in aiding malicious parties to build turnkey malware. 

The remainder of the paper is organized as follows. The next section briefly examines the prior art in this field. Section~\ref{S:Background} offers the necessary background information. Section~\ref{S:Testbed} details our testbed and methodology, while Section~\ref{S:Experiments} presents the results. The last section concludes and offers directions for future work.

\section{Related work}
\label{sec:related}

The current section briefly reviews the most closely related works to ours in the literature so far. The focus is on malware obfuscation and other evasion tactics, as well as contributions that experimentally assess the detection ability of AV and similar products. Therefore, works that examine different aspects, say, static and dynamic malware analysis techniques, are purposefully left out. The analysis is confined to the period between 2010 and 2023, considering both survey works as well as research contributions.

With respect to AV evasion, the authors in~\cite{Afianian2019} contributed a comprehensive survey on malware analysis, specifically on ways to evade dynamic analysis techniques for both manual and automated modes. In this regard, they first split the considered evasion tactics into two major categories, namely detection-dependent, and detection-independent. Secondly, they proposed a comprehensive classification of these techniques and showcased their performance vis-\'a-vis distinct detection and analysis schemes. Likewise,~\cite{Marpaung2012} surveyed malware evasion techniques, covering obfuscation (including encryption and polymorphism), fragmentation and session splicing, application-specific violations, protocol violations, inserting traffic at IDS, DoS, and code reuse attacks, such as return-oriented programming (ROP). In addition, they discuss mitigations and compare evasion techniques in terms of sophistication, detection difficulty, and impact. An interesting highlight from their analysis is that obfuscation is the only evasion technique with low sophistication and detection difficulty that has a higher-than-low impact; the rest of the techniques with medium or high impact have higher sophistication or detection difficulty requirements.
In the same context, the authors in~\cite{You2010} surveyed different malware obfuscation techniques. Their analysis focused on four types of malware: encrypted, oligomorphic, polymorphic, and metamorphic. For the last two techniques, various obfuscation schemes were discussed, including dead-code insertion, register reassignment, subroutine reordering, instruction substitution, code transposition, and code integration.

In addition, the work in~\cite{ntanto2019} exploited ROP for code obfuscation. Recall that ROP, basically a powerful code reuse technique, is used to counteract common exploit prevention schemes, including data execution prevention (DEP) and address space layout randomization (ASLR). Specifically, the authors highlighted two key contributions compared to the prior art. The first relates to the automatic analysis and generation of code snippets to produce relevant ROP chains, while the second pertains to the reuse of legitimate code extracted from executables for the purpose of creating ROP gadgets. The latter are short sequences of instructions ending in a ``ret'', that is, a return command in assembly that transfers control to the return address inserted in the stack by a ``call'' command. They introduced a tool to effectively transform a  piece of shellcode to its ROP equivalent. This is done by reusing the available code in the executable and then patching the ROP chain, finally infecting the executable. According to their results, this transformation results in an undetectable behavior when tested against the well-known VirusTotal scanning service.

Furthermore, the authors in~\cite{RIGAKI2023103192} centered on model stealing attacks devoted to malware detection. A model stealing or model extraction attack in the ML ecosystem works by querying the target model with samples and subsequently exploiting the model responses to duplicate it. Their study concentrated on standalone malware classifiers and AV products. Among others, they detailed a new model stealing attack, which combines transfer and active learning, and evaluated it through a series of relevant experiments. 

With respect to the benchmarking aspect, the authors in~\cite{Leka2022} examined the effectiveness of desktop versions of a dozen of popular AV products against their VirusTotal equivalents. Their dataset comprised 50 pieces of malware, which have been generated using 16 different open-source AV evasion tools that obfuscated two Metasploit payloads. According to their results, desktop AV software outperformed VirusTotal for most AV detection engines. The authors conclude that this could be attributed either to the lack of cloud-based detection on VirusTotal or to different configurations of VirusTotal and desktop AV engines.
Similarly but focused on mobile malware only, the work in~\cite{Rastogi2013} evaluated Android anti-malware software against common techniques used to obfuscate known and unknown malware. Their results showed that none of the 10 popular commercial anti-malware software tested was resistant to the utilized techniques, which included even slight transformations to known malware. This latest observation is in accordance with our results showing that legacy malware obfuscated through publicly and easily accessible techniques is still effective in AV evasion. The main difference is that we consider desktop AV engines, whereas~\cite{Rastogi2013} limit their study to Android anti-malware software only.

Based on the foregoing analysis, the most closely related work to the current one is that in~\cite{Leka2022}. Particularly, quite similar to~\cite{Leka2022}, we empirically assess the faculty of 16 popular commercial AV and EDR products in detecting legacy malware code, i.e., code that is publicly available, say, in GitHub repositories. Nevertheless, differently to~\cite{Leka2022}, the present study examines the effectiveness of different common and uncommon (but practicable) methods in obfuscating malicious pieces of code with the aim to pass unnoticed.

\section{Background}
\label{S:Background}

This section summarizes the necessary background regarding the malware detection and evasion techniques used in the context of this paper. In short, the basic techniques used by the tested AV and EDR products towards examining if a piece of code is malicious or not are two~\cite{Samociuk23}: signature-based and heuristic detection. On the other hand, malware authors utilize several methods to avoid detection; the methods used in the context of this paper are summarised in Table~\ref{T:techniques}.

The first detection technique typically relies on the detection of malicious patterns by only scanning the program or piece of code for known strings (or signatures). Other widely used methods for signature-based detection include fuzzy hashing (a function that compares similarities between files) and YARA rules, which are used to create descriptions of malware families based on textual or binary patterns. Essentially, the suspicious file is scanned for patterns of known header files, libraries, or even packed files. 

Heuristic analysis methods can be either static or dynamic, using machine learning or data mining techniques for effective malware detection. Static analysis involves decompiling a suspect program to inspect the source code and compare it to known viruses that reside in the heuristic database. In case a significant percentage of similarities is observed, the code is flagged as potentially malicious. Dynamic heuristics on the other hand, execute the suspicious piece of code inside an isolated, specialized VM, often called ``sandbox'', to test the code and decide about its functionality. Generally, dynamic analysis inspects the suspicious file during its execution and its runtime behavior is mapped to patterns of malicious activity. Generally, contrary to signature-based scanning, which looks to match signatures found in files with that of a database of known malware, heuristic scanning uses various decision rules or weighing methods.

On the flip side, malware writers exploit different methods towards hiding the malicious code and elude detection. Table~\ref{T:techniques} recaps some of the most prominent traditional AV evasion methods used in our analysis~\cite{Afianian2019}. The right column of the table designates the targeted type of malware code analysis considered in the context of this work for each evasion technique. One of the most prevalent methods to bypass malware analysis relies on the elimination of the prospect to match a malware part with a known signature. This can be easily achieved by hiding the code, say, by means of encryption. 
This technique entails splitting the malware into two pieces, the body, and the encryption/decryption function. The latter function is used to encrypt the body of the malware before its propagation. Then, when the malware is about to be executed, the function is used to decrypt the body of the malware which is then executed as usual.
Taking AES as an encryption/decryption function example, a variety of encryption modes of operation can be used, including AES-CBC, AES-CFB, and others.

\begin{table}[thb]
    \centering
    \begin{tabular}{|l|l|}
        \hline
        \textbf{Evasion technique}  &\textbf{Code analysis methods}  \\ \hline
        Encryption & Signature  \\
        \hline
        Process injection & Signature, Heuristic \\
        \hline
        Manual shellcode mem loading & Signature, Heuristic  \\
        \hline
        API hashing & Signature, Heuristic \\
        \hline
        Junk data & Signature  \\
        \hline
        Multiprocessing & Heuristic \\
        \hline
        Chosen shellcode & Heuristic \\        
        \hline
    \end{tabular}
    \caption{Malware evasion techniques used in the context of this work}
    \label{T:techniques}
\end{table}

With reference to Table~\ref{T:techniques}, process injection is another widespread defense evasion technique used to elude signature and heuristic analysis. In this case, the adversary injects pernicious chunks in the address space of a legitimate process; essentially, the malicious code is triggered when the legitimate process is executed. In this way, the malicious code running in the context of a legitimate process may enable access to the resources of the latter. Since the execution is masked under a legitimate process, the malicious code can stay under the radar of security products. For more information on process injection, the reader is referred to~\cite{hosseini2017ten} and~\cite{MITRE-PI}.
 
A third approach to overcome both types of malware analysis concerns the manual loading of a shellcode in the victim's memory. This, typically, involves a call to the \textit{VirtualAllocEx} or \textit{WriteProcessMemory} application programming interface (API). Specifically, this technique is used to defeat certain detection methods that rely on static analysis, since the code is only loaded at runtime and, therefore, is expected to be invisible to the underlying analysis system. Traditionally, if a piece of software needs to call a function of the Windows API, say, \textit{CreateFileW} to create a file, the software would need to reference the API name ``directly'' in the code. However, by doing so, the name of the API is left present in the code. This enables the defender to easily identify what the piece of suspicious code might be doing.

API hashing is another technique used by malware creators to overcome signature-based and heuristic analysis. With reference to the MS Windows OS, a piece of malware would typically populate the import address table (IAT) with references to API function names it uses, such as \textit{VirtualAlloc}, \textit{CreateFileW}, \textit{CreateThread}, and \textit{WaitForSingleObject}. The analysis of the IAT by a security analyst, however, could easily reveal what actions a suspicious piece of code performs. By using API hashing, an attacker can avoid this by obfuscating the function calls using hashing. In other words, the names of API functions are replaced with a hashed value, which generates a unique checksum for every file. The digests are used to carry out the API calls in a pseudonymous manner. This makes it harder for analysts and security products to expose the malevolent tasks performed by a malware.

Hiding the malicious code inside junk data is another popular old-school technique among malware creators. This involves appending junk data to the malware executable to inflate the file size in hopes of staying below the radar of, say, YARA rules.

Creating or injecting multiple processes with different shellcodes is yet another technique used to evade AV solutions. The main idea behind this concept is as follows. The malware author creates or injects multiple instances of a process, each with a different shellcode, into a target system. Then, the coder monitors antivirus detection and identifies which shellcode(s) has not been detected. Finally, the malicious payload can be executed in the undetected shellcode.

The last technique, namely chosen shellcode based on condition, is similar to the previous one but here the shellcodes are selected based on the target system's specifications and configuration. The first step in this case is to identify the target system and collect information, such as OS version and installed AV software. Then, the malware author develops multiple shellcodes based on the conditions of the target system and executes them to identify a shellcode that goes undetected by the AV.

It should be noted here that the combination of two or more techniques of Table~\ref{T:techniques} or the utilization of the same scheme multiple times, depending on the case, is also a viable approach.

\section{Testbed and methodology}
\label{S:Testbed}

As already pointed out, the goal of the present work is to assess the capacity of popular AV and EDR products in detecting legacy malware when obfuscated with the techniques of Table~\ref{T:techniques}. The experiments were performed on a Windows 11 Enterprise Edition VM equipped with 8 GB RAM and a quad-core CPU. Each one of the selected AV products was evaluated on a separate, clean VM instance, provisioned with the latest OS updates. After its installation, each AV had access to the Internet. The malicious executables were transferred over the Internet using public file transfer applications, like Filebin.

A dozen of the currently most popular AV products were tested, as presented in Table~\ref{T:AV}. We split the various products into free and trial/paid editions. As observed from the same table, we additionally evaluated four EDRs, specifically those which offer a free trial period. Note that each EDR suite incorporates one central cloud web application that allows the management of a set of connected (slave) devices. This caters for protecting the end-user of each one of the slave devices and alerting them when detecting a suspicious action. The version of each evaluated security product is mentioned in Table~\ref{T:AV}.

\begin{table}[htbp]
\centering
\begin{tabular}{ll}
\hline \textbf{Product name}             & \textbf{Version}      \\\hline 
\multicolumn{2}{c}{\textbf{AV}}                         \\\hline 
\multicolumn{2}{c}{Free editions}              \\\hline 
Avast                           & 230320-4     \\
AVG                             & 230327-12    \\
Avira                           & 1.1.84       \\
MS Defender                     & 1.385.1272.0 \\\hline 
\multicolumn{2}{c}{Trial/Paid editions}        \\\hline 
Webroot                         & 23.1         \\
Eset Smart Security Premium     & 16.0.26.0    \\
Bitdefender Total Security      & 26.0.34.145  \\
Kaspersky Small Office Security & 21.9.6.465   \\
Sophos Home                     & 4.3.0.5      \\
MalwareBytes                    & 4.5.24       \\
McAfee Total Security           & 19.21.167    \\
Norton                          & 22.23.1.21   \\\hline 
\multicolumn{2}{c}{\textbf{EDR solutions}}              \\\hline 
Bitdefender Gravity Zone        & 7.8.4.270    \\
Sophos Central                  & 2022.4.2.1   \\
ESET Protect Cloud              & 10.0.2045.0  \\
MS 365 Defender                 & N/A          \\\hline 
\end{tabular}
\caption{List of tested AV/EDR products as of 30 March 2023 (N/A: not applicable)}
\label{T:AV}
\end{table}

For the needs of this work, 16 malware variants were created, as recapitulated in Table~\ref{T:test} for easy reference. The table is split into three parts, namely Original, Modified, and ChatGPT, corresponding to the three stages described in the rest of this section. The corresponding variants are evaluated in subsections~\ref{SS:Original:Executables} to~\ref{SS:Modified-ChatGBT}, respectively.

Initially, three different malicious code snippets were built using three popular programming languages, namely C++, Go, and Rust. The latter two languages were purposefully chosen to evaluate the capacity of security products to detect malware written in modern languages. For all three code snippets, except our own code, we heavily reused ``as is'' publicly available malware code from GitHub~\cite{O:GO,O:API:Hashing,O:Rust}; this satisfies the objective of evaluating the capability of security products in detecting legacy malware. For reasons of reproducibility, all the three above-mentioned code snippets are available on a publicly accessible GitHub repository~\cite{O:bypass:repo}.

Furthermore, we exploited the Sliver\footnote{https://github.com/BishopFox/sliver} and NimPlant\footnote{https://github.com/chvancooten/NimPlant} C2 frameworks and Metasploit framework (MSF) to enable the attacker to acquire a reverse connection after the malware is executed. These three frameworks have been intentionally selected for the following reasons. First, Metasploit is the commonest penetration testing tool, and it is highly expected to be flagged by most AVs. Second, Sliver is a modern C2 solution that has been used in the wild, as a Cobalt Strike replacement~\cite{O:Sliver}; therefore, it is also anticipated to be flagged by AVs. Third, NimPlant is not yet a very popular lightweight first-stage C2 implant written in Nim, meaning that its detection may be more difficult for antiviral software.

Generally, such frameworks have the ability to generate different types of files, which if executed, would return a shell connection to the attacker. Regarding Sliver and NimPlant, such files (also known as \textit{implants}) can have the .dll, .exe, or .bin suffix. Among others, Metasploit can generate a hex string (the binary value of the string in hexadecimal notation) that the coder can then include in its own code written in a supported programming language, e.g., Go, Rust, and others. After that, if the compiled file is executed and the hex string is loaded into a process, the attacker will receive a shell connection. Note that this also requires the attacker to start a listener in the specific framework. For Sliver, we used the \textit{mTLS} listener, which establishes the default mutual TLS (authenticated) connection. For NimPlant and Metasploit, we exploited the default TCP listeners. The (default) beacon mode was used for Sliver and NimPlant and the session mode for Metasploit.

To cover different cases regarding the viral software, we chose the hex string method from Metasploit and the binary format (.bin) from Sliver and NimPlant. Precisely, through Sliver, we created a binary file, while for Metasploit we used the \textit{msfvenom} tool to generate a shellcode specifically for the Go language in the variants contained in the Modified part of Table~\ref{T:test}. Note that as shown in line 5 of Table~\ref{T:test}, the original Go code included a hex string that would pop up the Windows calculator, if the file is executed successfully. Additionally, the Go code included a manual way of allocating OS memory and executing the relevant process. Both the .bin file and the shellcode were encrypted by means of either AES-CBC or AES-CFB and imported to the original malware code snippets. Namely, the .bin files were imported to the C++ and Rust versions, and the shellcode to the Go snippet. The above-mentioned malware variants are included in the six topmost rows of Table~\ref{T:test}, where the first three rows refer to the original executable files as generated from each framework, without the use of any additional method.

\begin{table*}[htbp]
\centering
\begin{adjustbox}{width=1\textwidth}
\begin{tabular}{r|l|cccccccccccc} \hline
No &  Exe       & Enc & PI & APIH & MSF shell & MSF calc & Sliver & NimPlant & JD & Multi & Chosen & Manual & CS \\\hline
\multicolumn{14}{c}{Original}                                                                                   \\\hline
\rowcolor{lightgray}
1 & MSF      &            &                   &             & \cmark         &          &        &          &        &  & & &  \\
2 & Sliver   &            &                   &             &           &          & \cmark      &          &       &   & & &  \\
\rowcolor{lightgray}
3 & NimPlant &            &                   &             &           &          &        & \cmark        &         &   & & &\\
4 & C++      & \cmark          & \cmark                 & \cmark           &           &          & \cmark      &          &         &   & & &\\
\rowcolor{lightgray}
5 & Go       &            &                  &             &           & \cmark        &        &          &         &  & & \cmark  &\\
6 & Rust     & \cmark          & \cmark                 &             &           &          & \cmark      &          &         &   &  & &\\\hline
\multicolumn{14}{c}{Modified}                                                                                   \\\hline
\rowcolor{lightgray}
7 & C++      & \cmark          & \cmark                 & \cmark           &           &          & \cmark      &          & \cmark       &  & &  &\\
8 & C++      & \cmark          & \cmark                 & \cmark           &           &          & \cmark      & \cmark        & \cmark        & \cmark  & & &\\
\rowcolor{lightgray}
9 & Go       &  \cmark         &                  &             & \cmark         &          &        &          &        & & & \cmark &\\
10 & Go       &  \cmark         &                  &             & \cmark         &          &        &          & \cmark        & & & \cmark &\\
\rowcolor{lightgray}
11 & Rust     & \cmark          & \cmark                 &             &           &          & \cmark      &          &         &   & & &\\
12 & Rust     & \cmark          & \cmark                 &             &           &          & \cmark      &          & \cmark        &  & &  &\\
\rowcolor{lightgray}
13 & Rust     & \cmark          & \cmark                 &             &           &          &        & \cmark        & \cmark       &   & & &\\
14 & Rust     & \cmark          & \cmark                 &             &           &          & \cmark      & \cmark        & \cmark     &   & \cmark & &\\\hline 
\multicolumn{14}{c}{ChatGPT}                                                                                   \\\hline
\rowcolor{lightgray}
15 & ChatGPT      &           & \cmark                 &            &           &          &       &          &       &     & & &\cmark\\
16 & ChatGPT      &           & \cmark                 &            &           &          &       &          & \cmark &   & & &\cmark\\\hline
\end{tabular}
\end{adjustbox}
\caption{List of the 16 generated instances of malware (Enc: encryption, PI: process injection, APIH: API hashing, MSF: Metasploit framework, JD: junk data, Multi: multiprocessing, Chosen: chosen shellcode, Manual: manual execution of process in memory, CS: custom shell)}
\label{T:test}
\end{table*}

As a second stage, with reference to Table~\ref{T:test}, we altered the original code of C++, Go, and Rust, in order to make detection harder. These eight modifications are given in the mid-part of the table, under the ``Modified'' title. Precisely, the first variant of the original C++ code was modified to include junk binary data. Junk data were generated via the \textit{/dev/urandom}, which provides an interface to the Linux kernel's random number generator, allowing one to specify the size of the created file. This technique was also used for the Rust and Go code snippets, as shown in Table~\ref{T:test}. In all cases, the size of the produced random data file was 80 MB, except for variants destined to test the Avast and AVG products; in the latter case, the file size was 300 MB. This was done because, based on our preliminary tests, the Avast and AVG AV products flagged every executable file under 300 MB for further checking to their cloud services. Most probably, this behavior is on the grounds that our executables were unsigned. Nevertheless, via the simple trick of surpassing this file size threshold, we forced the Avast and AVG AVs to analyze our executables on the spot and not flag them for being unsigned, causing their cloud services to be invoked. Further, a second modification regarding the C++ code was to include both Sliver and NimPlant binaries, with the purpose to execute them at the same time and inject different processes.

For the Go executable, first, we added encryption/decryption, the MSF shellcode, and generated two versions, i.e., one with junk and another without junk data as shown in lines 9 and 10 of Table~\ref{T:test}. For the Rust instance, we slightly modified the original code, removing the sleep functions and the conditional execution, i.e., the executable in the original version had to be executed from a \textit{start} command with the \textit{activate} argument.

Next, as exhibited in lines 11 and 12 of Table~\ref{T:test}, two versions were created, with and without junk data. Moreover, as demonstrated in line 13 of the same table, we replaced Sliver binary with that of NimPlant and kept the junk data. The last Rust variant included both Sliver and NimPlant binaries along with junk data. For this variant, we observed that some AVs were able to detect the Sliver process. Therefore, to avoid detection, the code was instructed to check if the directory of one of these AV exists. If true, the executable loaded into the injected process the NimPlant binary.

As an additional step, ChatGPT was assessed in terms of producing effective malicious code. A first observation was that ChatGPT was unable to produce encryption/decryption functions correctly for Windows 11 APIs; human intervention was needed in all cases for correcting the generated code. For this reason, another approach was used, namely opening a listener to the victim and having the attacker connect to it. In this way, the encryption function is not needed, and in any case, it enabled the malware to masquerade as a legitimate process since many applications establish Internet connections in this way. As a result, ChatGPT code included only a process injection function, with and without junk data, and a shellcode generated by ChatGPT. This shell initiates a TCP connection between the attacker and the victim. The attacker sends a \textit{cmd} command, the victim retrieves the command, executes it, and sends the result back to the attacker.

All executables in Table~\ref{T:test} have been programmed for stealthy operation, preventing visual notifications, such as the cmd window pop-up. To treat each Sliver, NimPlant, or Metaspoit connection similarly, we waited for 5 min after its establishment. Then, we executed one command every 2 min, for a total of 10 min. Subsequently, we started all over again, waiting for a further 5 min period. If the antiviral product did not block the executable after three consecutive loops, command execution was ceased, classifying the corresponding variant as undetectable.

\section{Experiments}
\label{S:Experiments}

\subsection{Original executables}
\label{SS:Original:Executables}

As a first step, we evaluated the detection capacity of every AV product against each original malicious executable (.exe) file. 
The detection ratio per AV product is seen in the rightmost column of Table~\ref{T:first:results}. The results for MSF were according to our expectations, i.e., all AVs detected the malicious executable, even before a user managed to execute it. This means that in all cases, the static analyzer in each AV flagged it as malicious. However, this is not the case with both Sliver and NimPlant, where only half of the AV solutions managed to flag both executables as malicious. The detection rate was even worse for the other three executables:
only five out of 12 AVs managed to detect all three executables, whereas two AVs detected none of them. C++ and Rust had the best score, evading five out of the 12 AVs each. 

\begin{table*}[htbp]
\centering
\begin{adjustbox}{width=0.7\textwidth}
\begin{tabular}{l|llllll|ll}\hline
AV           & MSF                   & Sliver                & NimPlant                   & C++                                 & Go                                              & Rust             & Quar. & DR \\\hline
\multicolumn{9}{c}{Free edition}\\\hline
Avast        & \xmark* & \xmark & \xmark & \xmark  & \xmark & \xmark&  1/6   & 6/6    \\
AVG          & \xmark* & \xmark & \xmark & \xmark  & \xmark & \xmark&  1/6   & 6/6    \\
Avira        & \xmark* & \cmark & \xmark* & \xmark*  & \xmark* & \xmark* &  5/6   & 5/6    \\
MS Defender & \xmark* & \xmark* & \xmark* & \cmark  & \cmark & \cmark &  3/6   & 3/6    \\\hline
\multicolumn{9}{c}{Trial/Paid edition}\\\hline
Webroot      & \xmark* & \xmark & \xmark & \cmark  & \cmark & \xmark&  1/6   & 4/6    \\
Eset         & \xmark* & \cmark & \xmark* & \cmark  & \xmark* & \xmark&  3/6   & 4/6    \\
BitDefender  & \xmark* & \xmark* & \cmark & \xmark  & \cmark & \cmark&  2/6   & 3/6    \\
Kaspersky    & \xmark* & \xmark* & \cmark & \xmark*  & \xmark* & \cmark&  4/6   & 4/6    \\
Sophos       & \xmark* & \xmark & \cmark & \xmark  & \xmark & \xmark&  1/6   & 5/6    \\
MalwareBytes & \xmark* & \xmark* & \xmark* & \cmark  & \xmark & \cmark&  3/6   & 4/6    \\
McAfee       & \xmark* & \cmark & \xmark* & \cmark  & \cmark & \cmark&  2/6   & 2/6    \\
Norton       & \xmark* & \xmark* & \xmark* & \xmark*  & \xmark* & \xmark &  5/6   & 6/6    \\\hline
Total     & 0/12 & 3/12 & 3/12 & 5/12 & 4/12 & 5/12 & --   &  --\\\hline
\end{tabular}
\end{adjustbox}

\caption{Results regarding the original executables of Table~\ref{T:test}. The \xmark~and~\cmark symbols per AV line are seen from the attacker's perspective, standing for ``detect'' and ``evade'', respectively. The same symbols are used the same way in the rest of the tables of this section. The star superscript denotes a quarantined (after successful detection) executable file. (Quar.: quarantined file, DR: detection ratio)}
\label{T:first:results}
\end{table*}

Considering all the original executables, the comparison between free and paid AVs shows that the former scored better. Indeed, half of the free AVs identified all malware and only 17\% of the evasion attempts were successful. On the other hand, only 1 out of 8 paid AVs identified all malware, and the rate of successful evasions almost doubled, climbing to 33\%. In total, approximately 28\% of all executables managed to evade AV scanners; this corresponds to 20 successful evasions over 72 tests.

\subsection{Use of common evasion techniques}
\label{SS:Modified}

In this experiment, we used common evasion techniques, as described in Section~\ref{S:Background} and demonstrated in the ``Modified'' part of Table~\ref{T:test}. Such techniques include using uncommon programming languages for implementing existing malware, adding AES-CBC encryption to the Go executable~\cite{O:Go:AES}, and refactoring the Rust code to avoid signature detection. The corresponding results are presented in Table~\ref{T:sec:results}. Overall, in comparison with the original executables shown in Table~\ref{T:first:results}, we observed that the detection rate of both free and paid AVs dropped significantly when we employed simple malware modification techniques.

\begin{table*}[htbp]
\centering
\begin{adjustbox}{width=0.9\textwidth}
\begin{tabular}{l|llllllll|ll}\hline
AV           & Go                   & Rust                & C++ $\otimes$                 & Go $\otimes$          & Rust $\otimes$       & Rust $\omega$$\otimes$ & C++ $\diamond$$\otimes$ & Rust $\diamond$$\otimes$        & Quar. & DR \\\hline
\multicolumn{11}{c}{Free edition}\\\hline
Avast        & \xmark & \xmark & \cmark  & \xmark* & \cmark & \cmark  & \cmark & \cmark &  1/8   & 3/8    \\
AVG          & \xmark & \xmark & \cmark & \xmark* & \cmark & \cmark
& \cmark & \cmark &  1/8   & 3/8    \\
Avira        & \xmark* & \xmark* & \cmark & \cmark & \cmark & \cmark  & \cmark & \cmark &  2/8   & 2/8    \\
MS Defender & \cmark & \cmark & \cmark &  \cmark& \cmark & \cmark
& \cmark & \cmark &  0/8   & 0/8    \\\hline
\multicolumn{11}{c}{Trial/Paid edition}\\\hline
Webroot      & \cmark & \xmark & \cmark & \cmark & \cmark & \cmark
& \cmark & \cmark &  0/8   & 1/8    \\
Eset         & \xmark* & \cmark & \cmark &  \xmark*& \cmark & \cmark  & \cmark & \cmark &  2/8   & 2/8    \\
BitDefender  & \xmark & \cmark & \xmark & \xmark & \cmark & \xmark  & \xmark* & \cmark &  1/8   & 5/8    \\
Kaspersky    & \cmark & \cmark & \xmark* &  \cmark& \cmark & \cmark  & \xmark* & \cmark &  2/8   & 2/8    \\
Sophos       & \cmark & \xmark & \xmark &\cmark & \xmark & \cmark
& \cmark & \cmark &  0/8   & 3/8    \\
MalwareBytes & \cmark & \cmark & \cmark & \cmark & \cmark & \cmark
& \cmark & \cmark &  0/8   & 0/8    \\
McAfee       & \cmark & \cmark & \cmark & \cmark & \cmark & \cmark
& \cmark & \cmark &  0/8   & 0/8    \\
Norton       & \cmark & \xmark* & \cmark & \cmark & \cmark & \cmark
& \cmark & \cmark &  1/8   & 1/8    \\\hline
Total     & 7/12 & 6/12 & 9/12 & 8/12 & 11/12 & 11/12& 10/12& 12/12 & --   &  --\\\hline
\end{tabular}
\end{adjustbox}

\caption{Analysis of executables based on common evading techniques (Quar.: quarantined files, DR: detection rate, $\otimes$: includes randomly generated files, $\diamond$: includes both binaries, i.e., Sliver and NimPlant, $\omega$: includes NimPlant binary, $\star$: illustrates the binaries that were quarantined)}
\label{T:sec:results}
\end{table*}

Regarding the first technique, the use of an uncommon language, like Go or Rust, can increase the chances of evading AV software. Indeed, with reference to the first column of Table~\ref{T:sec:results} and compared to Table~\ref{T:first:results}, when combining Go with MSF shellcode the evasion rate increased from 33\% to 58\%. As seen in the second column of Table~\ref{T:sec:results}, refactoring the Rust code increased the evasion rate from 42 to 50\%, i.e., an increase of 19\%.

Another simple approach was based on augmenting the size of an executable file with randomized data; this also proved to be highly effective in evading AV solutions. To this end, we implemented a test scenario in which a malicious executable was enlarged by 100 MB using randomly generated data to prevent detection. Our results showed that the evasion rates of all executables vastly increased. In more detail, for C++ the evasion rate raised from 42 to 75\% (an increase of 79\%), for Go from 33 to 67\% (an increase of 100\%), and for Rust from 42 to 92\% (an increase of 120\%). In addition, we tested the Rust code with the NimPlant binary file; in this case, the results were the same as with Rust with only randomized data.

The next scenario involved loading both Sliver and NimPlant binary files on different processes, during the same execution. When using this technique with C++ we managed to evade an extra AV solution compared to using C++ with randomized data only, as shown in Table~\ref{T:sec:results}. Taking a closer look at this additional AV solution, we discovered that it was unable to detect and trace one of the two processes of the malicious executable. Using the same approach with Rust and combining it with the chosen shellcode based on condition technique, as described in Section~\ref{S:Background}, we managed to bypass all AV solutions. More specifically, the condition we used was to check which AV solution was installed in the target OS and utilize the binary that could evade this AV.

\subsection{ChatGPT}
\label{SS:Modified-ChatGBT}

Ever since the release of the popular ChatGPT tool, various malicious groups have attempted to exploit its capabilities to develop undetectable malware programs~\cite{O:ChatGPT:Malware}. For this reason, as already mentioned in Section~\ref{S:Testbed}, we investigated the possibility of using ChatGPT 3.5 (v. March 14) for developing malware with the C++ programming language, which is the most prevalent language for this purpose. Our aim was to make ChatGPT generate malicious code independently without any assistance in the coding process. To achieve this, we posed a question, received a response, tested the response using a compiler, and provided compile errors as feedback to the tool to rectify any errors.
In the end, this attempt was unsuccessful since ChatGPT had issues in generating a functional obfuscating code instance, such as encoding or encryption, for C++ and the latest native APIs of Windows 11.

Consequently, we adopted an alternative approach to generate the obfuscating code. We asked ChatGPT to produce a simple TCP listener, similar to an SSH listener, which would enable an attacker to connect and execute Command Prompt (cmd) commands, utilizing only native APIs of Windows. A precondition for making the listener operational was to have the firewall port required for connection open. Nevertheless, we considered this condition as beyond the scope of our paper, assuming that the user could already have this port open, since many legitimate programs, such as MS Teams, could request such access or be easily tricked to open it using other methods, such as social engineering.

Even though the usual approach would be for the victim to connect to a C2 server, the above unconventional method allowed us to evaluate the potential of ChatGPT to meet the same objective of executing malicious commands to a victim's system remotely. The first scenario in this case was for the attacker to establish a connection to the deployed malware using a standard connection tool, such as \textit{Netcat}, and generate various cmd commands; the corresponding detection rates for each AV are provided in Table~\ref{T:chatgbt:results}.

In the second scenario, we analyzed the executable produced by the code snippet provided by ChatGPT, and subsequently, we added a randomly generated 100 MB file and repeated the test. Similar to the previous subsections, adding junk data proved to be very effective in evading AVs; here, four additional AV solutions were evaded, which corresponds to a 79\% increase compared to the first scenario. Overall, only 3 out of 12 AVs managed to detect both ChatGPT scenarios, whereas on the other end, five detected none. In order to ease the reproducibility of our experiments, the relevant code by ChatGPT has been made publicly available in the main GitHub repository of this project~\cite{O:bypass:repo}.

\begin{table}[htbp]
\centering
\begin{adjustbox}{width=0.45\textwidth}
\begin{tabular}{l|ll|ll|r}\hline
AV           & ChatGPT & ChatGPT $\otimes$  & Quar & DR & \textbf{ODR}\\\hline
\multicolumn{5}{c}{Free editions}\\\hline
Avast        & \xmark & \xmark &  0/2   & 2/2  & \textbf{11/16}  \\
AVG          & \xmark & \xmark &  0/2   & 2/2  & \textbf{11/16}  \\
Avira        & \xmark* & \cmark &  1/2   & 1/2 & \textbf{8/16}   \\
MS Defender  & \cmark & \cmark &  0/2   & 0/2  & \textbf{3/16}  \\\hline
\multicolumn{5}{c}{Trial/Paid editions}\\\hline
Webroot      & \cmark & \cmark &  0/2   & 0/2   & \textbf{5/16} \\
Eset         & \xmark* & \cmark &  1/2   & 1/2  & \textbf{7/16}  \\
BitDefender  & \xmark* & \cmark &  1/2   & 1/2  & \textbf{9/16}  \\
Kaspersky    & \xmark* & \xmark* &  2/2   & 2/2 & \textbf{8/16}   \\
Sophos       & \cmark & \cmark &  0/2   & 0/2   & \textbf{8/16} \\
MalwareBytes & \cmark & \cmark &  0/2   & 0/2   & \textbf{4/16} \\
McAfee       & \cmark & \cmark &  0/2   & 0/2   & \textbf{2/16} \\
Norton       & \xmark* & \cmark &  1/2   & 1/2  & \textbf{8/16} \\\hline
Total      & 5/12   & 9/12   &   --   &  --  & --\\\hline
\end{tabular}
\end{adjustbox}
\caption{ChatGPT and final comparison (Quar: quarantined files, DR: detection rate, $\otimes$: including randomly generated file, $\star$: illustrates the binaries that were quarantined). The rightmost column presents the overall detection rate (ODR), calculated as the sum of the corresponding DRs included in Tables~\ref{T:first:results} to~\ref{T:chatgbt:results}}.
\label{T:chatgbt:results}
\end{table}

\subsection{EDR analysis}

To complement our study, we also scrutinized common EDR solutions that provided trial versions for evaluation purposes. Table~\ref{T:EDR} contains the respective detection rates of each executable when analyzed by each EDR product, namely ``Bitdefender Endpoint Security'', ``Sophos Endpoint Agent'', and ``ESET Endpoint Security''. The results indicate that the detection rates are similar to those of the corresponding AV solutions of the same vendor. At first glance, this suggests that EDR solutions, apart from improved administrative capabilities and monitoring of devices in an infrastructure, do not offer additional protection against the malicious executables we implemented using simple techniques. Consequently, such attacks pose a threat not only to home but also to enterprise users, regardless of their company's size.

\begin{table*}[htbp]
\centering
\begin{adjustbox}{width=1\textwidth}
\begin{tabular}{l|llllllllllllllll|ll}
\hline
EDR & MSF & Sliver & NimPlant & C++ & C++ $\otimes$ & C++ $\diamond$$\otimes$ & Go $\varpi$ & Go & Go $\otimes$ & Rust & Rust $\varpi$ & Rust $\otimes$ & Rust $\omega$$\otimes$& Rust $\diamond$$\otimes$ & ChatGTP & ChatGPT $\otimes$ & Quar & DR \\\hline
Bitdefender     &  \xmark*   &     \xmark*   &     \cmark    &   \xmark  &    \xmark      &     \xmark     &    \cmark   &  \xmark*  &     \xmark    &   \cmark&   \cmark   &      \cmark     &     \xmark      & \cmark&    \xmark*    &       \cmark       &    4/9  &  9/16  \\
Sophos          &  \xmark*   &     \xmark*   &     \cmark    &   \xmark  &    \xmark      &     \cmark     &     \xmark*  &  \cmark  &     \xmark    &   \xmark   &     \xmark      & \xmark      &     \cmark     &  \cmark&   \cmark    &       \cmark       &  3/9    &  9/16  \\
ESET            &  \xmark*   &   \xmark     &    \xmark*      &   \cmark  &     \cmark     &     \cmark     &    \xmark*   &  \xmark*  &     \xmark*    &   \cmark   &  \cmark   &   \cmark       &     \cmark      &  \cmark&   \cmark    &     \cmark         &   5/6   &  6/16  \\
MS 365 Defender &   \xmark*  &    \xmark*    &     \cmark     &  \cmark   &    \cmark      &      \cmark    &    \cmark   &  \xmark*  &     \cmark    &   \cmark   &  \cmark   &    \cmark      &    \cmark       &  \cmark&  \cmark     &        \cmark      &   3/3   &  3/16  \\\hline
Results         &  0/4   &    0/4    &     3/4     &   2/4  &      2/4    &    3/4      &   2/4    &   1/4 &   1/4      &   3/4   &     3/4  & 3/4     & 3/4  &   4/4       &    3/4     &      4/4        &    --  &   --\\\hline
\end{tabular}
\end{adjustbox}
\caption{EDR solutions evaluation ($\otimes$: includes randomly generated file, $\diamond$: includes both binaries, i.e., Sliver and NimPlant, $\varpi$: denotes that this is the original executable found in the relevant public GitHub repository, $\omega$: includes the NimPlant binary, $\star$: illustrates the binaries that were quarantined)}
\label{T:EDR}
\end{table*}

\subsection{Discussion}
\label{SS:Discussion}

To the best of our knowledge, two of the evasion techniques used in this paper are novel compared to existing literature: (i) manipulating multiple processes with various shellcodes and (ii) choosing a shellcode based on specific conditions (e.g., the existence of a folder). The operation of both of these techniques has been presented in Section~\ref{S:Background}.

Overall, our analysis showed that common AV/EDR solutions can be circumvented by employing simple, well-known techniques. Firstly, robust encryption algorithms can provide protection against static analysis conducted by most AVs. Secondly, the use of less common compiled programming languages with Windows native APIs can further assist in evading AV detection. Additionally, the use of common C2 servers, such as Sliver and NimPlant, can provide an advantage in the evasion process. Furthermore, incorporating a randomly generated file into the compiled executable demonstrated a significant improvement in evasion rates against all AV/EDR solutions. It should be noted here that the final executable that managed to evade all twelve (12) AV solutions was developed with Rust, employed AES-CTR encryption, included randomly generated files in the code, and checked which AV was running in the target system before initiating the process with the binary that could evade detection by AV/EDR.

Regarding the use of randomly generated files, this technique increased the size of each executable file to over 30-40 MB on average, which prevented a malicious file from being uploaded for cloud analysis. Interestingly, Avast and AVG considered a C++ executable file suspicious if its size was less than 300 MB. To test this, we created two different executable files, one with a size of 120 MB and the other with a size of 320 MB, to observe how these two AV/EDR solutions would behave. When the first file was executed, Avast and AVG blocked it as suspicious and attempted to upload it to the cloud, while the second file executed normally. Furthermore, if this executable file lacked the ``Mark of the Web (MOTW)'' flag, Avast, AVG, and MS Defender did not block its execution. Overall, Avast and AVG marked almost all executable files as malicious, possibly because they were not signed with a certificate.
Nevertheless, an attacker could easily circumvent this by either signing the file with a public certificate or using the popular DLL side-loading technique.

The observation of the behavior of different AV solutions during testing leads to some interesting remarks. With MS Defender, we observed unusual behavior when executing the Meterpreter process; more specifically, the process was flagged as malicious if we attempted to execute commands immediately after receiving the shell. To work around this, we waited a few minutes before executing commands, and the heuristic detection of MS Defender was then unable to detect the malicious Meterpreter process. Avira and Kaspersky on the other hand, were able to quarantine any initially flagged malicious executable, achieving a 100\% quarantine rate. Other AV solutions could detect the malicious processes but failed to connect them back to the original executable, which left the malicious file unquarantined and potentially harmful. Webroot was the only AV solution that worked in collaboration with MS Defender, which may explain why their detection rates were similar. Kaspersky had a unique behavior of asking the user to scan the entire operating system every time a malicious file or process was found, which could be quite burdensome for the user. However, Kaspersky was the only AV solution that correctly detected and stopped the malicious process of the ChatGPT executable. Finally, when analyzing the ChatGPT executable with Avast or AVG, we observed that as soon as the executable requested firewall access, these AV solutions flagged it as malicious. This is an indication that these free AVs may use generic methods to flag malicious files, which could lead to false positives and flagging legitimate files as malicious.

The last column of Table~\ref{T:chatgbt:results} provides a comprehensive overview of the effectiveness of all AV solutions against the various malware variants developed in this work. These results demonstrate that 42\% of the AV solutions were found to detect less than 50\% of the malicious executables, four (33\%) were able to detect exactly half of the malicious executables, while the remaining three (25\%) detected more than half of the malicious executables. It is worth noting here that none of the AVs detected all malicious executables, whereas the best score was achieved by two free AVs and it was equal to 11/16, i.e., 69\% detection rate. On the paid AVs, the best score achieved was 9/16, resulting in a detection rate of 56\%. On the other end, there were AVs scoring as low as 2, 3 and 4 out of 16, i.e., between 13 and 25\%; furthermore, the detected executables belonged exclusively to the original executables category, that is, the topmost part of Table~\ref{T:test}.

We found no disparity in the detection rates of AV and EDR software from the same vendor. This suggests that EDR editions provide the same level of protection as the relevant AVs and the only additional features are of administrative nature, i.e., to aid in the monitoring and management phases of an organization. Possibly, to increase the detection rate of an EDR solution, this should be properly configured, based on the needs of the relevant network infrastructure.

\section{Conclusions}
\label{sec:conclusions}

Antivirus software is a key weapon in the quiver of defenders in the never-ending fight against malware. This work investigates the detection performance of popular AV products in finding malicious pieces of code. Contrary to the existing literature, our main aim is to assess this capacity in cases where the malware strain contains legacy malicious code, but exploits uncommon evasion methods to a greater or lesser extent. Altogether, with reference to Table~\ref{T:techniques}, we utilize seven malware evasion techniques, applying them either individually or in tandem on 16 malware instances. Notably, to the best of our knowledge, three of these techniques, namely Junk data, Multiprocessing, and Chosen shellcode have not been evaluated so far against standard commercial AV products. No less important, we make an initial assessment regarding whether the ChatGPT chatbot platform can be tricked into generating malicious code that can slip through undetected. The obtained results came largely as a surprise, showing that only two of the AV engines (including EDRs which revealed identical detection scores with the AV product of the same vendor) were capable of detecting 11 out of the 16 malware variants. This result suggests that the majority of AVs may fail to detect legacy malware if it is concealed using not-so-common or uncommon but still ordinary techniques, such as encrypting the malware body, employing less popular programming languages, adding junk (random) data in the executable, and other similar methods. In this respect, AVs may provide an incorrect perception of security, not only lagging behind the latest and zero-day threats but also behind inventive tricks and techniques conjured by malware writers. Moreover, this study ends up with a set of key observations regarding the behavior of particular AV products when facing malware. For instance, while some AVs did discern the malicious processes, they backfired, not isolating the executable. A clear direction for future work is to expand this study by testing a richer repertoire of malware variants against a larger set of antivirus engines.


\clearpage
\bibliographystyle{ACM-Reference-Format}
\bibliography{sample-authordraft}

\appendix


\end{document}